# Development of an energy-sensitive detector for the Atom Probe Tomography

Christian Bacchi[1], Gérald Da Costa[1], Emmanuel Cadel[1], Fabien Cuvilly[1], Jonathan Houard[1], Charly Vaudolon[1], Antoine Normand[1] and François Vurpillot[1]

[1]Normandie Université, UNIROUEN, INSA Rouen, CNRS, Groupe de Physique des Matériaux, 76000 Rouen, France

**Abstract:** A position-energy-sensitive detector has been developed for APT instruments in order to deal with some mass peak overlap issues encountered in APT experiments. Through this new type of detector, quantitative and qualitative improvements could be considered for critical materials introducing mass peak overlaps, such as nitrogen and silicon in TiSiN systems, or titanium and carbon in cemented carbide materials.

This new detector is based on a thin carbon foil positioned on the front panel of a conventional MCP-DLD detector. According to several studies, it has been demonstrated that the impact of ions on thin carbon foils has the effect of generating a number of transmitted and reflected secondary electrons that mainly depends on both the kinetic energy and the mass of incident particles. Despite the fact that this phenomenon is well known and has been widely discussed for decades, no studies have been performed to date for using it as a mean to discriminate particles energy. Therefore, this study introduces the first experiments on a potential new generation of APT detectors that would be able to resolve mass peak overlaps through the energy-sensitivity of thin carbon foils.

# I. Introduction

Nanotechnology is one of the fastest growing areas of research in materials science. The reason for that originates from the growing need to create new materials that would imply significant technological breakthroughs over the next decades (Miodownik, 2015). Whether it is a question of increasing the efficiency of solar cells, developing carbon storage at large scale or finding alternative to rare earth, it becomes essential to understand the internal nanostructures and microstructures of materials relying on those future technologies. With the aim of correctly characterizing nano-features, it is sometimes necessary to both determine their shape and their composition at the atomic scale. It turns out that the best analytical tool suited for those requirements is the Atom Probe Tomography (APT). Contrary to electron microscopes, APT instruments are able to both localize and chemically identify atoms, not only from samples surface, but also from internal layers of analyzed materials.

The basic principle of this instrument relies on the field evaporation of atoms from the material surface. By applying a high electric potential $V_{DC}$ (˜ 1 – 20kV) on a material sample shaped as a sharp tip, with a radius of curvature in the range 10 – 100 nm, it is possible to extract atoms from the sample surface. This phenomenon called “Field Ion Emission” (Tsong, 1990; Vurpillot, 2016), can be controlled by electric (or laser) pulses $V_P$, allowing atoms from the tip surface to be ionized and repelled towards a position-sensitive detector. During their projection, ions reach a kinetic energy $E_K$ that tends towards the value of the potential energy $E_P$ generated on the tip apex.

$$E_K = \frac{mv^2}{2} \quad (1)$$

$$E_P = ne(V_{DC} + V_P) \quad (2)$$

With m the mass, v the velocity and n the charge state of ion projectiles, and e the elementary charge of the electron. The theoretical conservation between those two energies (Equations 1 and 2) allows to determine a mass-to-charge ratio for each detected ion (Equation 3).

$$\frac{m}{n} = 2e(V_{DC} + V_P)\frac{TOF^2}{LOF^2} \quad (3)$$

To determine those mass-to-charge ratios, it is then necessary to perform time-of-flight (TOF) measurements between the start pulse, applied on the tip sample, and ions arrival on the detector. Therefore, this technique called “Time-of-Flight Mass Spectrometry” (TOF-MS), can provide information on the elemental nature of each detected ion. A 3D reconstruction of the analysed material is then computed with a back-projection algorithm, using both the order of arrival of each ion on the detector and the theoretical areal density of the sample surface (Bas et al., 1995; Gault et al., 2011; Hatzoglou et al., 2019).

Regarding the working process of the instrument, it might be thought that the APT could be the best mean for characterizing materials nanostructure with high spatial and mass resolutions. However, since its creation more than 30 years ago (Cerezo et al., 1988), the instrument is still not fully recognized as a reliable tool perfectly suited for material analyses. The first cause to this originates from the multiple difficulties to get reliable 3D reconstructions which sometimes do not really correspond to the reality (Miller, 1987; Vurpillot et al., 2000). The second main cause originates from biases brought by APT detectors.

One of the main biases brought by APT detectors relates to their inability to distinguish some elements one another. Indeed, it can be noticed that the composition of some analyzed materials may involve the evaporation of elements having almost equal mass-to-charge ratios. That is the case, for instance for nitrogen and silicon in TiSiN systems (Engberg et al., 2018) and in field-effect transistors (Martin et al., 2018), or titanium and carbon in cemented carbide materials (Thuvander et al., 2011). To overcome this limitation, it is necessary to couple the TOF-MS with a new spectroscopy technique that would be able to resolve those interferences, called mass peak overlaps (MPO).

It can be noticed that a significant part of MPOs concerns elements having both different charge states and different masses (La Fontaine et al., 2016; Engberg et al., 2018; Thuvander et al., 2011; Kirchhofer et al., 2014). Consequently, it can be deduced that a significant part of ions involved in MPOs could be indirectly resolved through the measurement of their kinetic energy (Equations 1 and 2). This assumption has already been stated few years ago (Kelly, 2011; Broderick et al., 2013), but has never been practically realized. This can be explained by the difficulty to measure additional physical quantities, while preserving the existing high performances of current APT instruments (Da Costa et al., 2012). Recent developments of position-energy-sensitive detectors (PESD) greatly illustrates this difficulty (Ohkubo et al., 2014; Fujii et al., 2015; Buhr et al., 2010; Funsten et al., 2004), where it can be stated that existing arrays of energy-sensitive cells introduce very slow response time, that does not go below hundreds of nanoseconds, conversely to few nanoseconds required in APT experiments. This limitation should have the effect of restricting the detection rate and the multi-hit capability of APT instruments. In addition to this, the limited number of energy-sensitive cells of those PESDs should also have the effect of degrading the high spatial resolution of current APT detectors. In cases the number of energy-sensitive cells may be increased, there is also the risk of increasing the complexity of the detection system through an extension of the parallel processing.

Other devices, that are to date not known or even used as PESDs, could both bring the lack of energy-sensitivity to APT instruments, and resolve those last limitations from existing PESDs. This applies to the secondary electron emission (SEE) induced by ion impacts on thin foils. It has been known for many decades that the secondary electron yield (SEY) generated from positive ion bombardment on thin foils is a function of both mass and kinetic energy of ion projectiles (Jackson, 1927; Oliphant, 1930; Hill et al., 1939). Until now the main practical purpose of thin foils was to be used as electron multipliers for getting start and stop signals for TOF spectrometers by being coupled with microchannel plates (Šaro et al., 1996a; Gloeckler & Hsieh, 1979; Shapira et al., 2000; Montagnoli et al., 2005). All this without considering the energy-sensitivity of thin-foils. Since most of those experiments were conducted with ion beams rather than through atom-by-atom analyses, it cannot be stated that this phenomenon could be useful for APT experiments. Moreover, very few publications have reported measurements of secondary electron yield (SEY) induced by ions, having the kinetic energy range encountered in APT experiments (around 5 keV – 40 keV). Therefore, the following study is both aimed at conducting experiments that could confirm the mass and energy-sensitivity of thin foils in the energy range of APT experiments, and at developing the first prototype of PESD for APT instruments, that would able to discretely resolve MPOs.

# II. Materials and Methods

## II.1. The Ion-Induced Secondary Electron Emission phenomenon

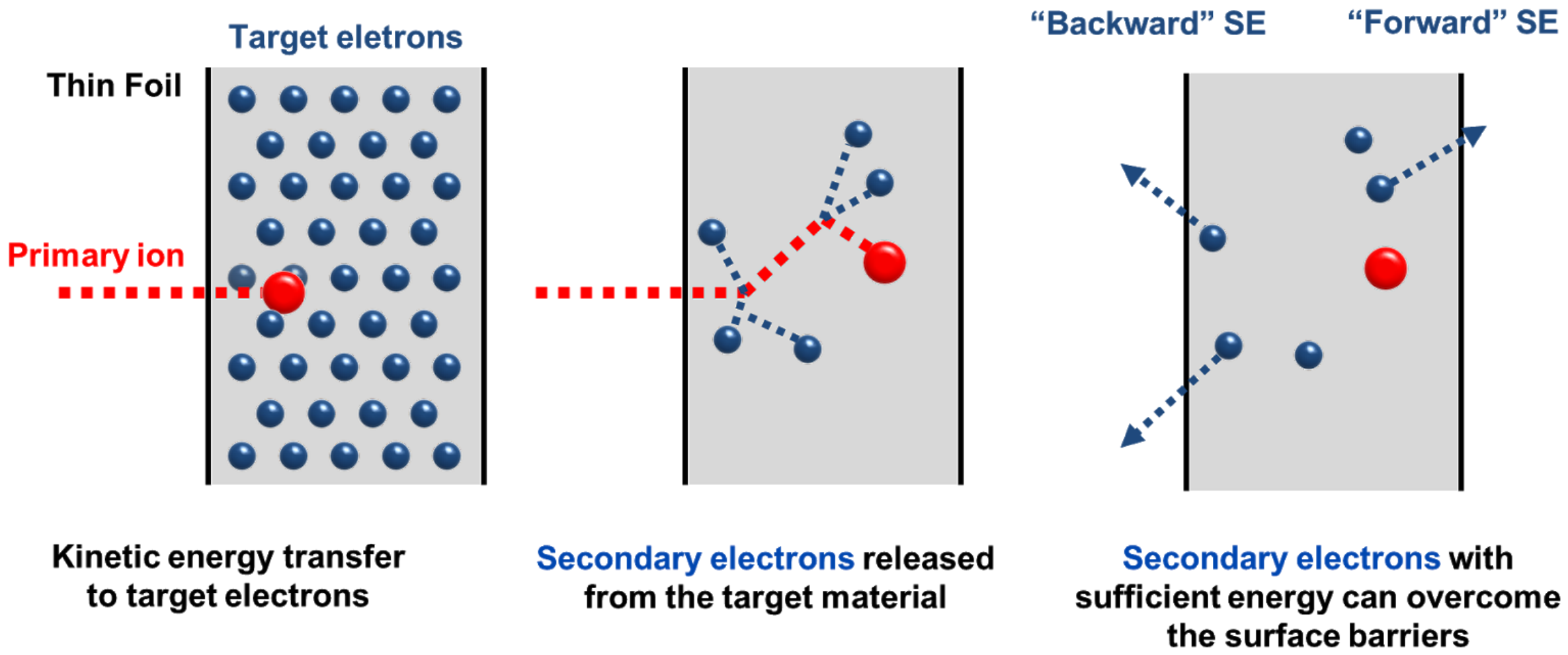

*Figure 1: Working principle of the Ion Induced Electron Emission on a thin foil. The penetration of the ion projectile inside the solid induces the generation of secondary electrons, which can be ejected from the entrance or the exit surface of the thin foil, as a function of their energy and their trajectory in the matter.*

The secondary electron emission (SEE) induced by ion-matter interactions, also known as Ion Induced Electron Emission, is generally considered as a three-step process (Rothard et al., 1995; Baragiola, 1993a) (Figure 1). First, an ion projectile transfers its kinetic energy to electrons from a target material; next, in case of thin foils, a fraction of these secondary electrons (SE) moves from the internal structure of the target material both towards the entrance and exit surfaces; finally, a fraction of those SE will be ejected from the target, provided that their kinetic energy is sufficient to overcome the surface barrier between the target and the vacuum medium.

It has been reported that most of these SE have an initial energy in the range 30 eV - 90 eV (Töglhofer et al., 1993; Baragiola, 1993a), then reduced to less than 5 eV for SE ejected to the vacuum medium (Palmberg, 1967; Rymzhanov et al., 2015; Allegrini et al., 2016; Dennison et al., 2006; Tomita et al., 2006). Moreover, studies have shown that those ranges of energies from SE do not depend on the nature of ion projectiles, nor on their kinetic energy, but mainly depend on the target material (Baragiola, 1993b; Hasselkamp, 1992). As for the photoelectric effect, some electrons from the target material are ejected after absorbing a specific part of the energy of the incident particle, exceeding the binding energy between those target electrons. At this point, it can be understood that the kinetic energy of an incoming ion will be discreetly distributed to a certain amount of target electrons around the point of impact. The more incoming ions are energetic, the more will be the amount of SE that could be released outside the thin foil.

## II.2. Collection of secondary electrons from carbon foils

By recalling that SE can be generated both from entrance and exit surfaces of thin foils (Figure 1), one can note that those SE can be collected at the proximity of those two surfaces. For that reason, different collecting techniques were developed for decades as a function of the different SEE applications (Šaro et al., 1996b). Most of those detection techniques use amorphous thin carbon foils (CF) in combination with Microchannel Plate (MCP) assembly for both getting timing pulses (start and stop time measurements), triggered by the SE, and for determining the original position of ion impacts (Shapira et al., 2000).

Regarding the different studies implying SEE from thin foils, there is evidence to suggest that carbon foils (CF) would be the best choice for generating high amounts of SE. Indeed, in 1955, Murdock et al. reported that CFs introduce one of the highest SEY compared with most other materials (Murdock & Miller, 1995). Although, there are exceptions such as cesium iodide (CsI) thin foils, introducing much higher SEYs than CFs (Chianelli et al., 1988), but with difficulties to manufacture robust and regular thicknesses under 100 µm (Chen et al., 2015; Weathers & Tsang, 1996). In counterpart, P. Maier-Komor (Maier-Komor, 1993) reported that CFs can be manufactured one order of magnitude thinner than foils made with other elements, with thicknesses below 5 nm. This can explain why most of current applications using the SEE phenomenon are using CFs (Allegrini et al., 2016). With the aim of correctly selecting the best CF thickness, specifically intended for APT analyses with ions energy range around 5 keV – 40 keV, one can refer to the synthetic study from Allegrini et al. (Allegrini et al., 2003), whose compared CF thicknesses from 0.5 µg/cm² to 10 µg/cm² (2.5 nm to 50 nm thick). From this study, it appears that CF thicknesses between 2 and 4 µg/cm² (approximately around 10 nm to 20 nm thick) are the most appropriate for getting the highest SEY in the energy range of APT experiments. This range of thicknesses can also be theoretically explained through a simple model considering both the mean distance that primary ions travel before generating SE, and the mean distance that SE travel inside the CF (Bacchi, 2020).

CFs have already been applied in several detection systems requiring ultra-high vacuum conditions in the same order of magnitude as APT instruments (< $10^{-9}$ Torr) (Kuznetsov et al., 2000; Drexler & DuBois, 1996). Moreover, reports from different space programs allow to calculate that CFs have a sufficient lifetime for performing more than one billion APT analyses without noticing any major difference in its SEE properties (Allegrini et al., 2016). From data originating from Allegrini et al. (Allegrini et al., 2016), it can be calculated that a CF surface of 40 $cm^2$ (~7 cm of diameter) would only lose less than 1 nm, by collecting ~ $8.10^{16}$ ions. For APT analyses comprising 100 million of ions each, this thickness loss would be reached after 800 million analyses.

In the framework of our study, particular interest has been paid to a configuration using the transitory secondary electrons (forward SE), with the aim of keeping a standard APT geometry, having a detection surface orthogonal to the probed sample (Figure 2).

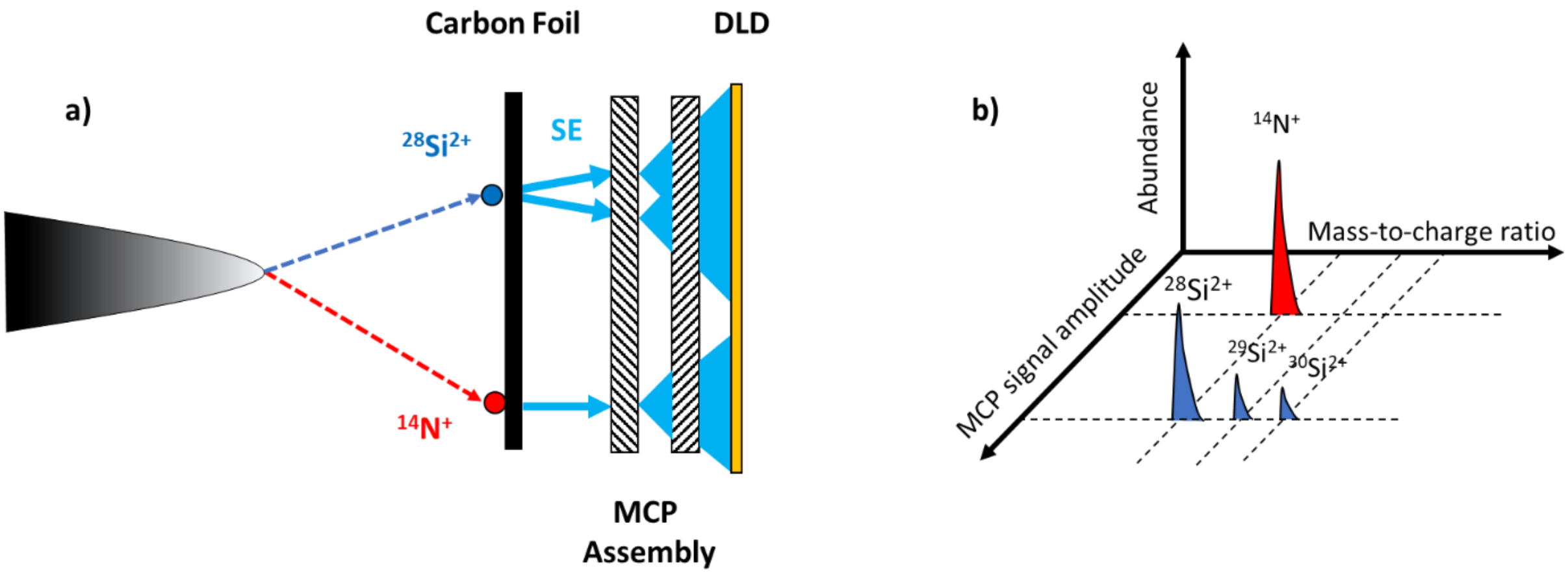

*Figure 2: Illustration of a material analysis performed with an APT instrument equipped with a CF detector; a) Schematic of the CF-MCP-DLD setup with the detection of $^{28}Si^{2+}$ and $^{14}N^{+}$, that conventionally leads to a MPO issue; b) Hypothetical mass spectra that would be obtained after the energy discrimination performed by the CF detector.*

## II.3. Measurement of the SEY induced by ion beams on carbon foils

In order to ensure the mass and energy sensitivity of CFs, preliminary studies have been conducted with the help of controlled ion beam sources, also known as Focused Ion Beams (Ward et al., 1985; Wortmann et al., 2013). Through a fine selection of ion projectiles and kinetic energies, it becomes possible to finely control the input parameters allowing the estimation of the SEY induced from CFs.

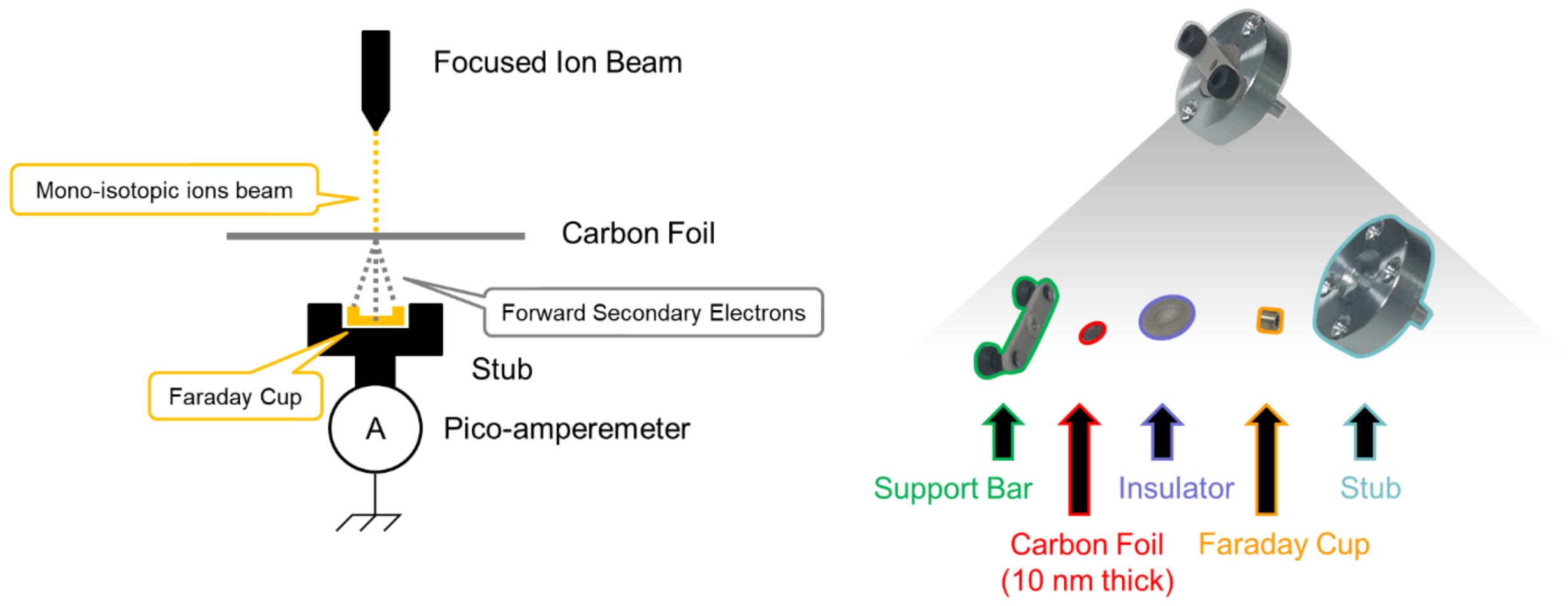

*Figure 3: Setup specially designed for SEY measurements on FIB instrument.*

Figure 3 introduces the device specially designed for those measurements. SEY measurements have been performed by collecting forward SE emerging from commercially available CF mounted on TEM grid from EMS (www.emsdiasum.com). The forward SEY has been determined by comparing the ion beam current $I_{beam}$, originating from the FIB source, to the current induced by forward SE $I_{SE}$ measured at the output of the foil. $I_{beam}$ and $I_{SE}$ currents have been simply measured through the same faraday cup and by means of a pico-ammeter (Figure 3).

According to the literature and the previous assessments, it must be recalled that ion projectiles may pass through the foil depending on their energy and their mass (Allegrini et al., 2003). It means that the current $I_{pA}$ measured on the pico-ammeter is the combination of a fraction $\alpha$ of incoming ions coming from the FIB source and forward SEs induced by ion impacts. Therefore, in order to get a reliable estimation of the SEY, special attention has been taken in the subtraction of the positive current induced by ions transmitted through the foil. In the framework of this this setup, currents induced by transmitted ions have only been statistically determined through the use of SRIM simulation tool (Ziegler et al., 2010; Ziegler & Biersack, 1985). The SEY can be then calculated through the following equation;

$$SEY = \left| \frac{I_{pA} - I_{beam} \times \alpha}{I_{beam}} \right| \quad (5)$$

Where $\alpha$ is the transmission coefficient of the foil calculated on SRIM.

It is also necessary to specify that, in order to avoid the electrostatic charging of the foil surface, its front end has been discharged through a tungsten micro-manipulator connected to an external ground.

SEY measurements were performed with two different beam sources, $Ga^{+}$ beam from a ZEISS - NVision 40 (www.zeiss.fr), and $Xe^{+}$ beam from a Helios G4 PFIB UXe DualBeam (www.thermofisher.com). Given that the respective masses of gallium and xenon are different nearly by a factor two (~131 Da for xenon and ~70 Da for gallium), it could be possible to reproduce a case of MPO through the analysis of $Xe^{+}$ ions having twice the kinetic energy of Ga+ ions. With the aim of ensuring high SEY from those heavy ions, CFs with a thickness of 2 µg/cm² (10 nm thick) has been chosen. Figure 4 clearly shows that SEY measurements from the two different ion beam sources ($Ga^{+}$ and $Xe^{+}$) provide a first confirmation that the number of forward SE really both depends on the mass and the kinetic energy of incident ions.

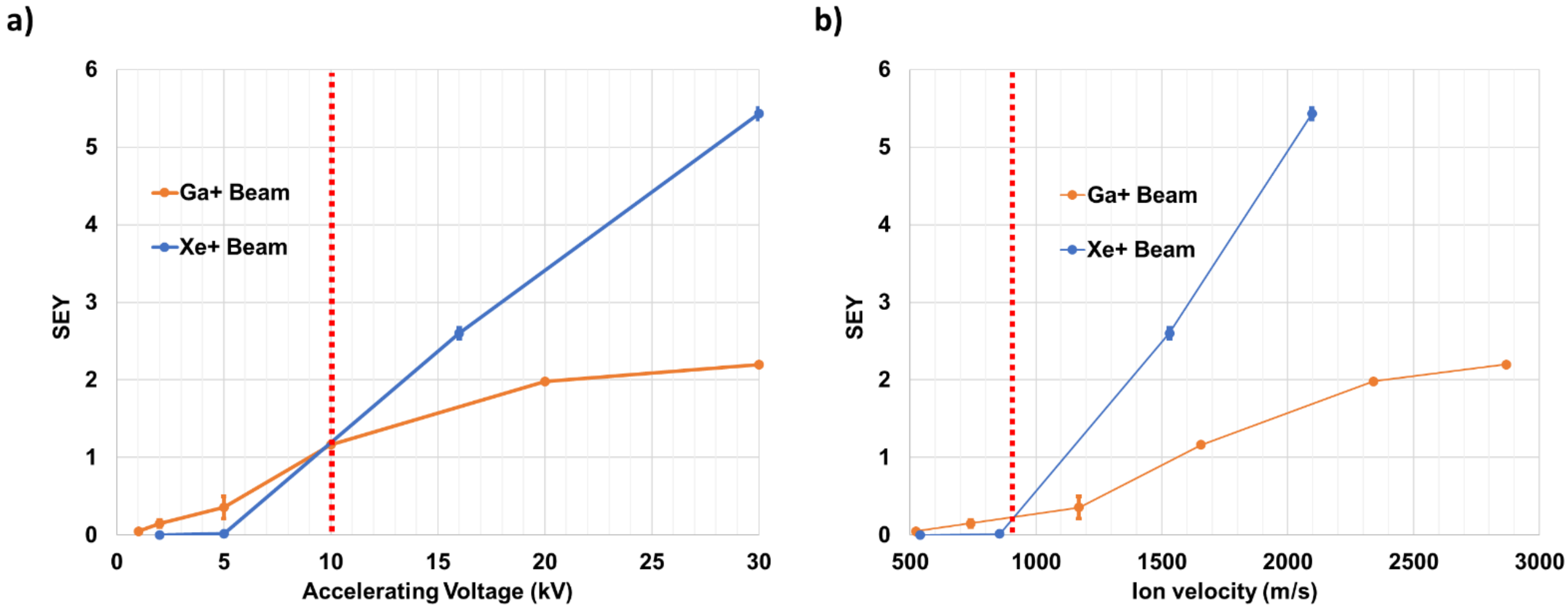

*Figure 4: SEY measurements from Ga+ and Xe+ ion beams as a function of the applied accelerating voltage (a), and as a function of ions velocity. The dotted red*

*lines represent the thresholds where Ga+ and Xe+ ions are always detected (SEY ≥ 1).*

From one side, the energy sensitivity of the CF is highlighted through the evolution of the SEY as a function of the accelerating voltage applied to ion projectiles (Figure 4a); from the other side, the mass sensitivity of the CF is highlighted through the evolution of the SEY as a function of ions velocity (Figure 4b). Specifying that, in APT experiments, very close mass-to-charge ratios also represent very close ion velocities. In this study, it can be observed that, from a specific velocity threshold, $Ga^+$ and $Xe^+$ ions with equal velocities introduce different SEY, with an increasing difference as a function of increasing velocities. Therefore, assuming that $Xe^+$ beam is set with twice the kinetic energy of the $Ga^+$ beam, it can be assumed that $Xe^{2+}$ and $Ga^+$ ions, having very close mass-to-charge ratios, could be distinguished through this technique. From this last observation, it becomes clear that MPOs implying elements with different charge states could be resolved through the use of CFs.

On the fact that those results only concern a statistical point of view of the phenomenon, it would be necessary to check if this method can be applied for single particle detection, such as for APT experiments.

## II.4. Single particle sensitivity of the Carbon Foil detector

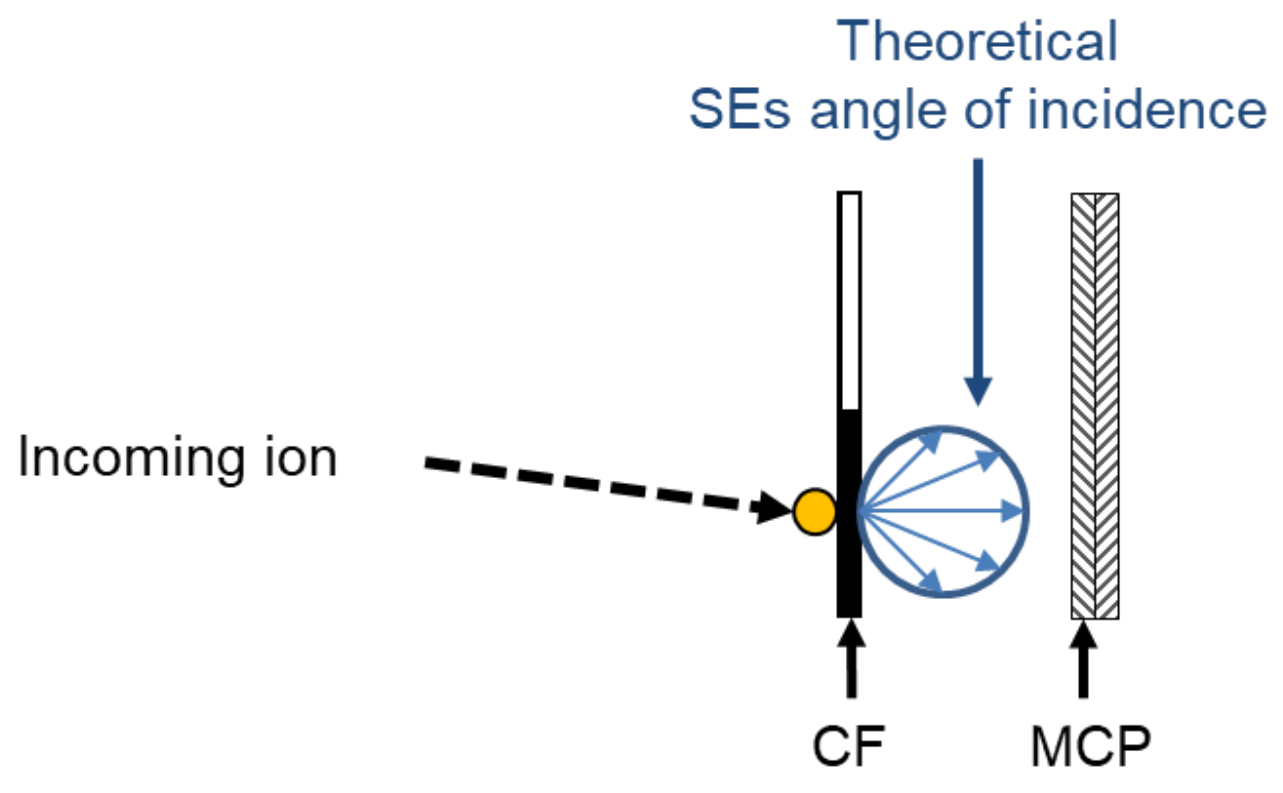

*Figure 5: Schematic representation of the angle of incidence of forward SE induced by an ion impact on a CF. Theoretically, the angular distribution of forward SE exhibits a cosine-dependence* (Hasselkamp, 1992; Ritzau & Baragiola, 1998).

According to the literature, it has been observed that the angular distribution of forward SE exhibits a cosine-dependence (Hasselkamp, 1992; Ritzau & Baragiola, 1998), suggesting that the distribution of the emitted SE is isotropic within the target. In that way, it would be highly probable that single ion hits would be transformed into unwanted SE multi-hits spread on the detection surface (Figure 5).

As a consequence, those potential SE multi-hits have to be spatially focused on a small area in order to get a single position and a single TOF for each single ion impact on the CF. To achieve that, it can be recalled that each impact treated with an MCP-DLD detector is surrounded by a dead-zone (DZ), where no other successive impact can be spatially resolved (Bacchi et al., 2019; Meisenkothen et al., 2015; Miller & Forbes, 2014; Peng et al., 2018; Jagutzki et al., 2002). Therefore, SE coming from a single ion impact on the CF must be focused on a small area that does not exceed this specific DZ. According to the study of Bacchi et al. (Bacchi et al., 2019), it is known that no successive impacts can be resolved within those following conditions;

$$\Delta TOF + DT \geq \frac{2|\Delta X|}{v_X} \geq \Delta TOF - DT \quad (5)$$

$$\Delta TOF + DT \geq \frac{2|\Delta Y|}{v_Y} \geq \Delta TOF - DT \quad (6)$$

Where $\Delta TOF$ is the TOF difference between ion pairs; $v_X$ and $v_Y$ are the transversal propagation velocities of electric signals respectively on X and Y delay line; $\Delta X$ and $\Delta Y$ are respectively the relative distances between multi-hit impacts on X and Y delay line axes; and DT is the instrument dead-time. Given that $\Delta TOF$ is nearly equal to zero for SE coming from a single ion impact on the CF, it can be deduced that the DZ where they must be confined together is reduced to the following conditions;

$$|\Delta X| \leq DT \frac{v_X}{2} \quad (7)$$

$$|\Delta Y| \leq DT \frac{v_Y}{2} \quad (8)$$

The combination of physical and electrical properties of our dedicated DLD with the dead-time our detection system (1.2 ns) made it possible to estimate a theoretical DZ of less than 2.99 mm × 1.77 mm.

In order to focus the transmitted SE on this restricted DZ, it is necessary to apply an electric field between the foil and the MCP front-end, for getting a single TOF and a single position for each single ion collection. To do so, a first electrostatic model has been defined for determining the optimal conditions for getting the best single particle sensitivity on a CF-MCP-DLD setup (Figure 6).

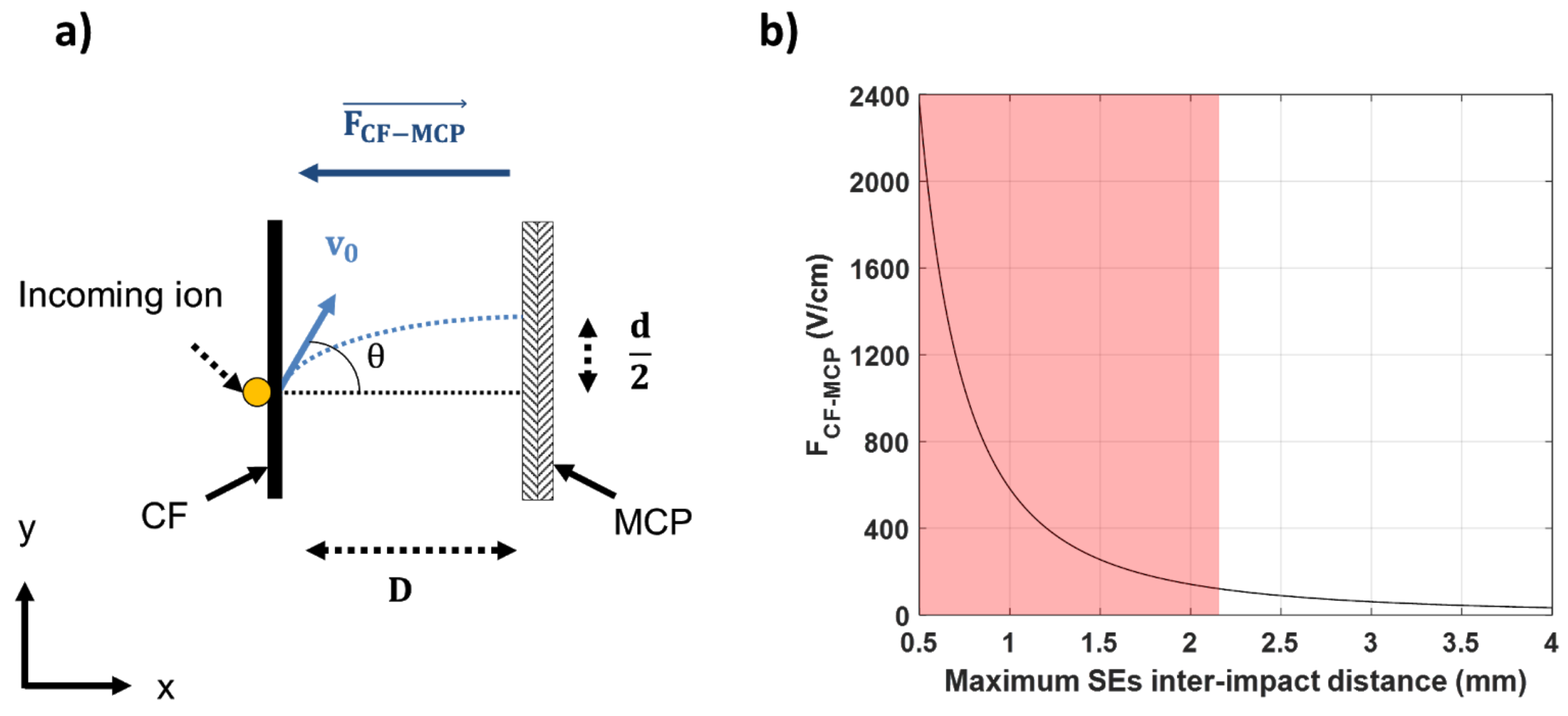

*Figure 6: Electrostatic model developed for focusing SE into a restricted DZ on the MCP front-end; a) Two-dimensional representation of the electrostatic model used for computing the required electric field for focusing SEs on a restricted area on the MCP; b) Graphical representation of the electrostatic model used for computing the required electric field for focusing SE on the restricted DZ. The red area defines the required electric fields for ensuring the focusing of SE in a same DZ.*

This model relies on three initial parameters;

- Due to mechanical constraints, the distance D between the foil and the MCP front-end could not be set under 1 cm. Thus, for maximizing the possibility to focus SE in very small areas, it has been decided to keep this distance to 1 cm.
- By considering the cosine dependence of the SE angular distribution (Hasselkamp, 1992; Ritzau & Baragiola, 1998), a mean angle of incidence ±60° related to the surface normal of the CF exit surface, has been set for covering half of the maximum probabilities of the angular distribution.

- The initial kinetic energy of SE has been set to a mean of 2 eV, in accordance with studies which showed that this order of magnitude does not depend on the kinetic energy or the mass of incoming ions (Allegrini et al., 2016; Tomita et al., 2006).

The graphical representation of this model in Figure 6b shows that the intensity of $F_{CF\text{-}MCP}$ must be higher than 123 V/cm for ensuring the focusing of SE in a same DZ, where the maximum inter-impact distance required is 2.14 mm on this setup.

Theoretically, it could be assumed that applying a very high value of $F_{CF\text{-}MCP}$ would be favorable for the single sensitivity of the CF detector, but two main limitations from this setup may induce significant losses. First, it should be noted that the saturation regime of the MCP assembly not only implies a limit of output charge from a single microchannel, but also from the microchannels that are very close to original points of impact (Ladislas Wiza, 1979; Barnstedt, 2016). This means that the very high proximity of SE can trigger the saturation of the MCP around adjacent microchannels. The second main limitation concerns the restricted electron detection efficiency of MCPs. It has been reported that the highest detection efficiency for electron projectiles can only be reached with energies varying between 200 and 400 eV (Goruganthu & Wilson, 1984; Müller et al., 1986). Given that the initial kinetic energy of SE outcoming from CFs is limited to few eV, it can be deduced that, for a distance D between the foil and the MCP front-end of 1 cm, $F_{CF\text{-}MCP}$ should be set between 200 V/cm and 400 V/cm.

## II.5. Experimental setup for the Carbon Foil detector

### II.5.1. CF-MCP-DLD assembly

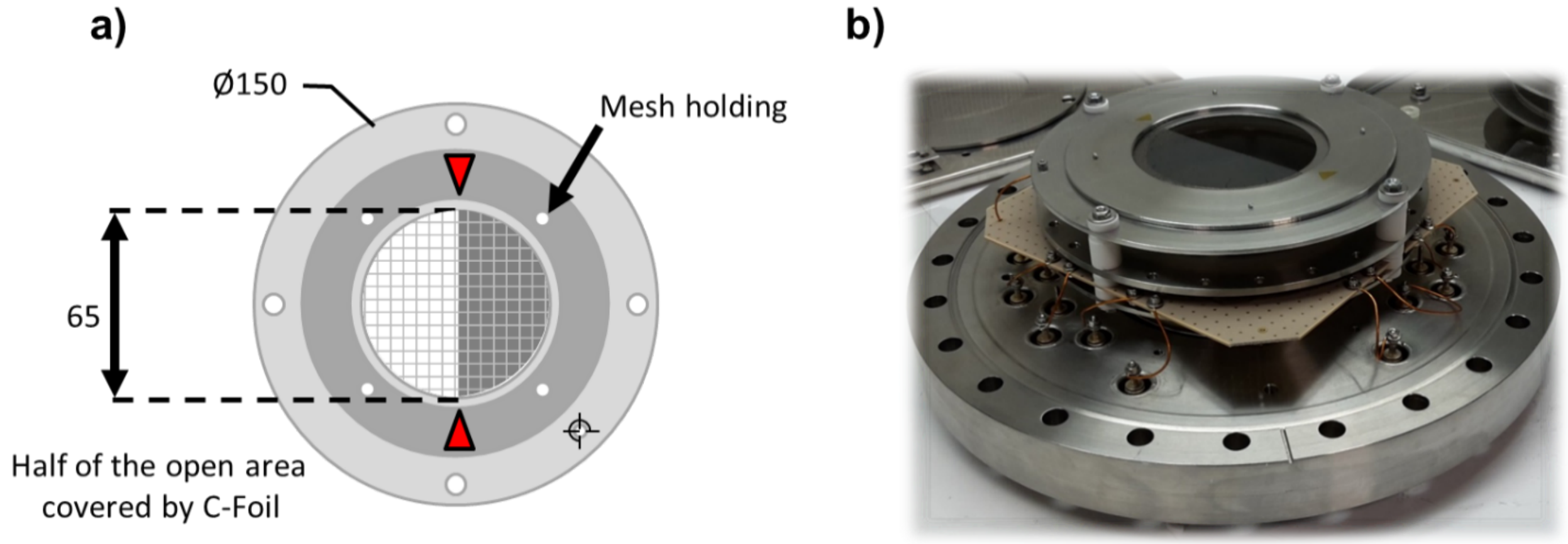

*Figure 7: Design of the Carbon Foil detector especially designed for this study. (a) Top view of the CF assembly introducing a detection surface half covered by a CF mounted on a 70 lpi Nickel mesh. (b) Photograph of the CF-MCP-DLD assembly mounted on a 200-CF flange.*

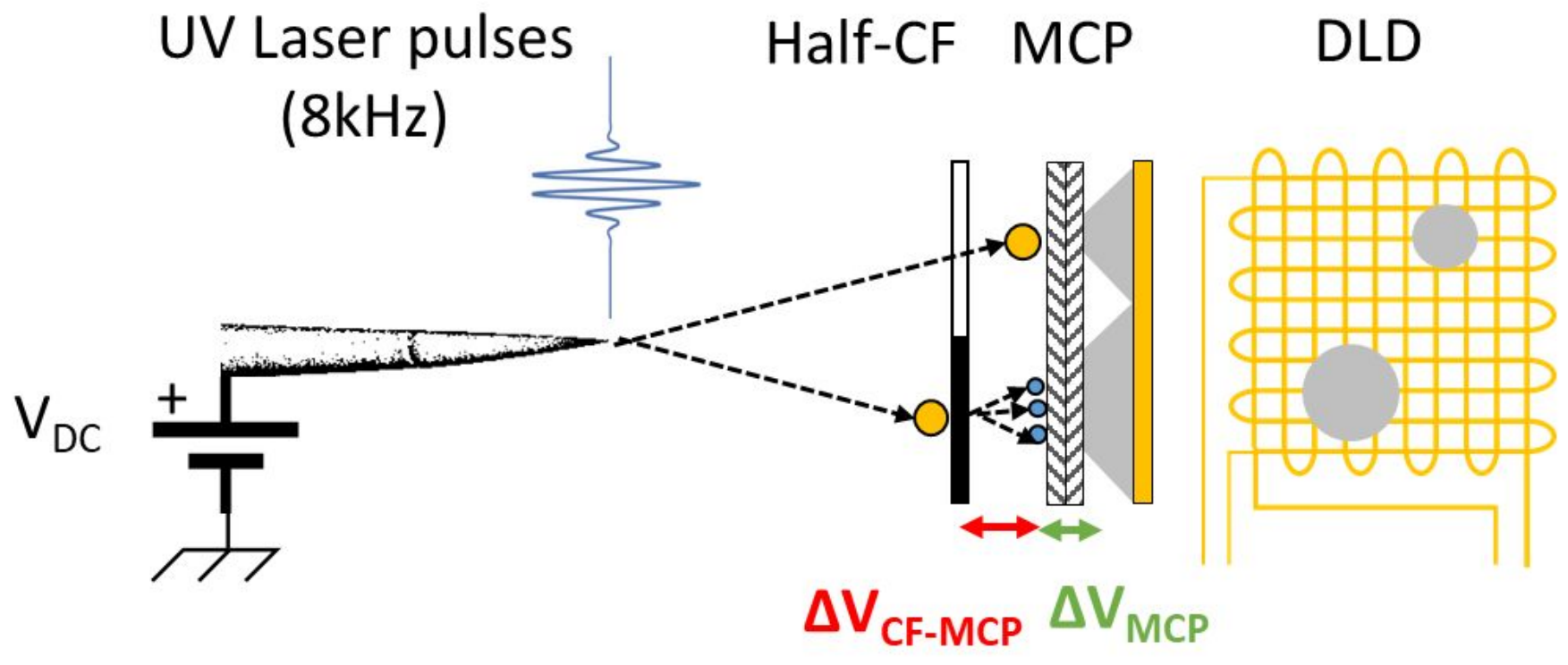

*Figure 8: Setup of the dedicated CF detector workbench.*

As previously mentioned, with the aim of fully exploiting the benefits of CFs mass and energy-sensitivity, it has been decided to combine a conventional MCP-DLD detector with a carbon foil mounting placed on its front-end (Figure 7). CFs of 2µg/cm² (~10 nm) thick were provided by Arizona Carbon Foil Company (ACF) and the mounting was designed at the GPM lab. In addition to providing information from the CF detector, a direct comparison with the conventional MCP-DLD detector has been performed by splitting the detection area in two distinct parts (Figures 7 - 8). The MCP assembly is characterized by an effective diameter of 77 mm, and an open area ratio of ~60% (nearly equal to the detection efficiency of the MCP assembly). The DLD used in this setup has been specifically designed at the GPM lab (Bacchi, 2020).

### II.5.2. Acquisition system of the Carbon Foil detector

In order to extract information from ions kinetic energy, it has been decided to use an APT detection system that is able to record amplitudes of output signals, known as the Advanced Delay Line Detector (Da Costa et al., 2005, 2012), or aDLD detection system. The aDLD detection system is based on fast digitizers, sampling MCP output signals at 4 Gs/s and DLD output signals at 1 Gs/s.

Indeed, in cases where single ion impacts respectively induce simultaneous SE multi-hits, it can be assumed that resulting MCP output signals will be highly dependent on the number of collected SE. As a consequence, the study of the MCP pulse height distribution can be used as a SEY indicator for all ion impacts.

# III. Results and discussion

## III.1. Analysis of a Fe40Al alloy

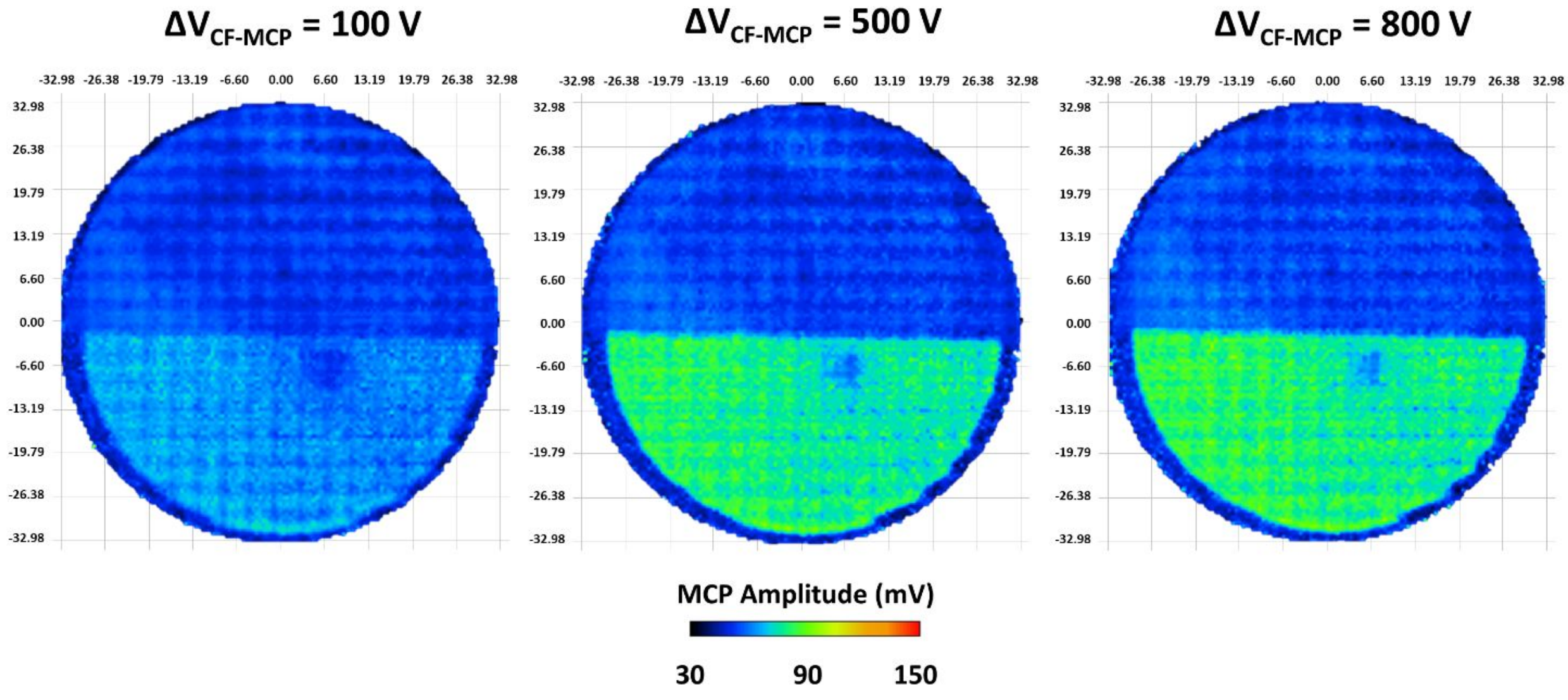

*Figure 9: 2D distribution of MCP amplitudes on the detector surface. The upper part of the detection surface is the area without foil and the lower part is the area covered by a CF. This figure clearly reveals the amplification of MCP signals through the CF*

*detector that is induced by the secondary electron emission for each ion impact on the foil.*

With the aim of evaluating the CF detector, a first analysis was performed with a Fe-40Al (at. %) alloy for highlighting the MPO located at 27 Da between the $^{27}Al^{+}$ peak and the $^{54}Fe^{2+}$ peak (Marceau et al., 2015; Seol et al., 2013). For this analysis, a DC voltage in the range 8.8 – 9.3 kV has been applied to the sample. In a global aspect, it can be primary observed in Figure 9 that the CF properly plays its role of signal amplifier. As discussed earlier, it can be observed that the increase of the electric field between the CF and the MCP has the effect of focusing SEs and creating overlapped electric signals for each ion impact. On the lower part of 2D maps (Figure 9), representing the CF area, it can be observed that MCP amplitudes are nearly twice higher than the upper part, representing the MCP-DLD area (without CF). It has to be specified that the narrow circular band of lower MCP amplitudes, following the edge of the CF area, is caused by the focusing of peripheral SE towards the center of the area through the applied electric field between the CF and the MCP front-end.

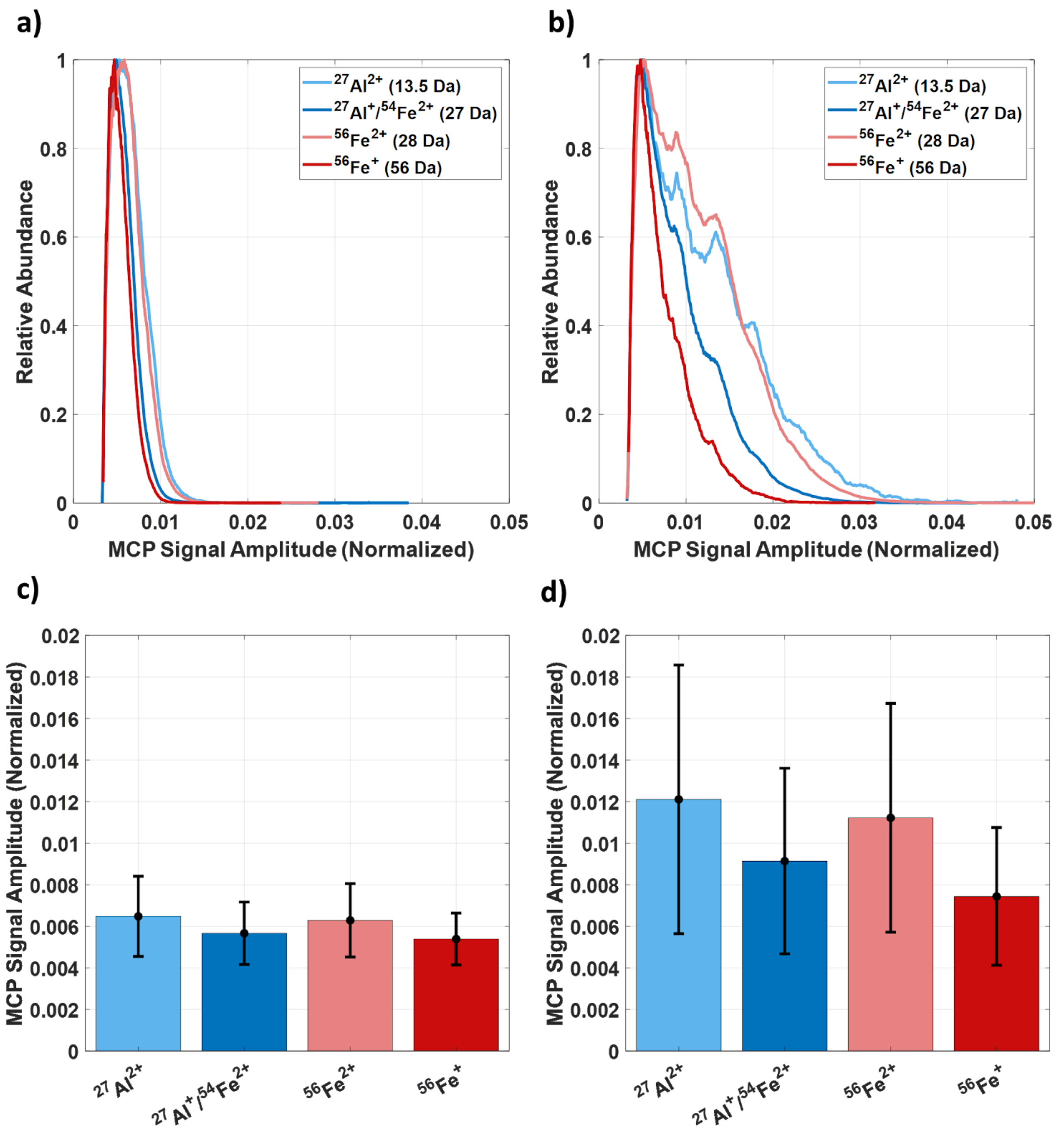

*Figure 10: MCP pulse height distributions originating from the analysis of the Fe40Al alloy: a) and c) from "no-foil" area, b) and d) from CF area (at $\Delta V_{CF\text{-}MCP}$ = 500 V). Histograms from the CF area (b and) show that detected ions with multiple charge state introduce higher amplitudes than ions with single charge state.*

Now concerning the potential MPO at 27 Da, a comparison between the two areas has been performed through their respective MCP pulse height distribution (Figure 10). Since there is a theoretical relation between the kinetic energy $E_K$ of ions and the electric potential V applied on the sample (Equations 1 and 2), MCP amplitudes have been normalized by each corresponding DC potential from each detection event. The MCP pulse height distribution from the CF area (Figures 10b and 10d) clearly shows that detected ions with multiple charge state introduce higher amplitudes than ions with single charge state. Which confirm again the energy-sensitivity of CFs.

| | $\Delta V_{CF\text{-}MCP}$ = 100V | | | | | |
|---|---|---|---|---|---|---|
| | No Foil (%) | Foil (%) | Gain/Loss (%) | Fe/Al Fraction (%) (No Foil) | Fe/Al Fraction (%) (Foil) | Gain/Loss (%) |
| $^{54}Fe^{+}$ | 1.55 | 0.56 | -0.99 | | | |
| $^{56}Fe^{+}/^{57}Fe^{+}$ | 24.75 | 6.50 | -18.25 | 63.75 | 62.09 | -1.66 |
| $^{56}Fe^{2+}/^{57}Fe^{2+}$ | 37.45 | 55.03 | 17.58 | | | |
| $^{27}Al^{+}/^{54}Fe^{2+}$ | 34.62 | 35.26 | 0.64 | 36.25 | 37.91 | 1.66 |
| $^{27}Al^{2+}$ | 1.63 | 2.66 | 1.02 | | | |

| | $\Delta V_{CF\text{-}MCP}$ = 500V | | | | | |
|---|---|---|---|---|---|---|
| | No Foil (%) | Foil (%) | Gain/Loss (%) | Fe/Al Fraction (%) (No Foil) | Fe/Al Fraction (%) (Foil) | Gain/Loss (%) |
| $^{54}Fe^{+}$ | 1.26 | 0.85 | -0.41 | | | |
| $^{56}Fe^{+}/^{57}Fe^{+}$ | 18.77 | 8.98 | -9.78 | 63.29 | 62.47 | -0.82 |
| $^{56}Fe^{2+}/^{57}Fe^{2+}$ | 43.26 | 52.64 | 9.37 | | | |
| $^{27}Al^{+}/^{54}Fe^{2+}$ | 33.05 | 33.40 | 0.35 | 36.71 | 37.53 | 0.82 |
| $^{27}Al^{2+}$ | 3.66 | 4.14 | 0.48 | | | |

| | $\Delta V_{CF\text{-}MCP}$ = 800V | | | | | |
|---|---|---|---|---|---|---|
| | No Foil (%) | Foil (%) | Gain/Loss (%) | Fe/Al Fraction (%) (No Foil) | Fe/Al Fraction (%) (Foil) | Gain/Loss (%) |
| $^{54}Fe^{+}$ | 1.00 | 0.93 | -0.06 | | | |
| $^{56}Fe^{+}/^{57}Fe^{+}$ | 16.97 | 9.80 | -7.16 | 63.27 | 62.32 | -0.96 |
| $^{56}Fe^{2+}/^{57}Fe^{2+}$ | 45.31 | 51.58 | 6.27 | | | |
| $^{27}Al^{+}/^{54}Fe^{2+}$ | 32.14 | 32.73 | 0.59 | 36.73 | 37.68 | 0.96 |
| $^{27}Al^{2+}$ | 4.58 | 4.95 | 0.37 | | | |

*Table 1: Comparison of the fractions of Al and Fe, in the analyzed Fe-Al alloy, between the MCP-DLD assembly and the CF-MCP-DLD assembly (Figures 7 - 8). Results from the different tables have been obtained from the same sample and at different bias voltage between the CF and the MCP front-end.*

At this point, it can be observed that the energy resolution obtained with this setup is not sufficient to totally resolve the MPO between $^{27}Al^{+}$ and the $^{54}Fe^{2+}$ mass peaks (Figure 10b). However, the significant difference on MCP amplitudes between ions with different charge states could still allow the spatial identification of elements that was not distinguishable in conventional APT instruments. Unfortunately, the homogenous nature of the analyzed Fe-Al alloy did not allow the clear visualization of the spatial separation between $^{27}Al^{+}$ and $^{54}Fe^{2+}$ ions. Therefore, other materials introducing MPOs in correlation with heterostructures must be analyzed for spatially revealing the energy sensitivity of the CF detector (see Analysis of an ZnO/$Mg_xZn_{1-x}O$ multi-quantum well system).

| | $\Delta V_{CF-MCP}$ = 100V | | |
|---|---|---|---|
| | Count (No Foil) | Count (Foil) | Relative Efficiency |
| $^{54}Fe^{+}$ | 16919 | 3040 | 17.97% |
| $^{56}Fe^{+}/^{57}Fe^{+}$ | 269586 | 35185 | 13.05% |
| $^{56}Fe^{2+}/^{57}Fe^{2+}$ | 407905 | 298044 | 73.07% |
| $^{27}Al^{+}/^{54}Fe^{2+}$ | 377101 | 190968 | 50.64% |
| $^{27}Al^{2+}$ | 17780 | 14383 | 80.89% |
| Total | 1089291 | 541620 | 49.72% |

| | $\Delta V_{CF-MCP}$ = 500V | | |
|---|---|---|---|
| | Count (No Foil) | Count (Foil) | Relative Efficiency |
| $^{54}Fe^{+}$ | 5947 | 3651 | 61.39% |
| $^{56}Fe^{+}/^{57}Fe^{+}$ | 88539 | 38633 | 43.63% |
| $^{56}Fe^{2+}/^{57}Fe^{2+}$ | 204093 | 226377 | 110.92% |
| $^{27}Al^{+}/^{54}Fe^{2+}$ | 155916 | 143627 | 92.12% |
| $^{27}Al^{2+}$ | 17265 | 17785 | 103.01% |
| Total | 471760 | 430073 | 91.16% |

| | $\Delta V_{CF-MCP}$ = 800V | | |
|---|---|---|---|
| | Count (No Foil) | Count (Foil) | Relative Efficiency |
| $^{54}Fe^{+}$ | 3979 | 3929 | 98.74% |
| $^{56}Fe^{+}/^{57}Fe^{+}$ | 67673 | 41277 | 60.99% |
| $^{56}Fe^{2+}/^{57}Fe^{2+}$ | 180702 | 217178 | 120.19% |
| $^{27}Al^{+}/^{54}Fe^{2+}$ | 128198 | 137808 | 107.50% |
| $^{27}Al^{2+}$ | 18270 | 20858 | 114.17% |
| Total | 398822 | 421050 | 105.57% |

*Table 2: Relative detection efficiencies of Al and Fe ions in the analyzed Fe-Al alloy, determined through the comparison between the amount of ions detected with the CF part, related to ions detected on the conventional MCP-DLD part (Figures 7 - 8). Results from the different tables have been obtained from the same sample and at different bias voltage between the CF and the MCP front-end.*

On the quantitative aspect of this study, one can ask whether the composition of the analyzed material has been altered by the ion-to-electron conversion performed by the CF. Since the design of the detector used in this study both includes a conventional MCP-DLD assembly and a CF-MCP-DLD assembly (Figures 7 - 8), it is possible to get an estimation of compositional biases brought by the CF detector, through a compositional reference brought by the conventional MCP-DLD assembly. Through this direct comparison (Table 1), it can be observed that low compositional biases are obtained within the required bias voltages $\Delta V_{CF\text{-}MCP}$ previously calculated. Those results can be partly understood by recalling that the highest MCP detection efficiency for electron projectiles can only be reached with energies varying between 200 and 400 eV (Goruganthu & Wilson, 1984; Müller et al., 1986).

Moreover, assuming that the SEY may exceed more than one SE per incoming ion (Figure 4), it can be assumed that probabilities to detect ions become higher than for conventional APT detection systems. Results from Table 2 confirm this assumption, where it can be observed that the detection efficiency of some elements can reach an increase of ~20%. It turns out that most of elements concerned by this increase are double charge ions; those expected to induce higher SEY related to single charge ions.

## III.2. Analysis of an ZnO/MgxZn1-xO multi-quantum well system

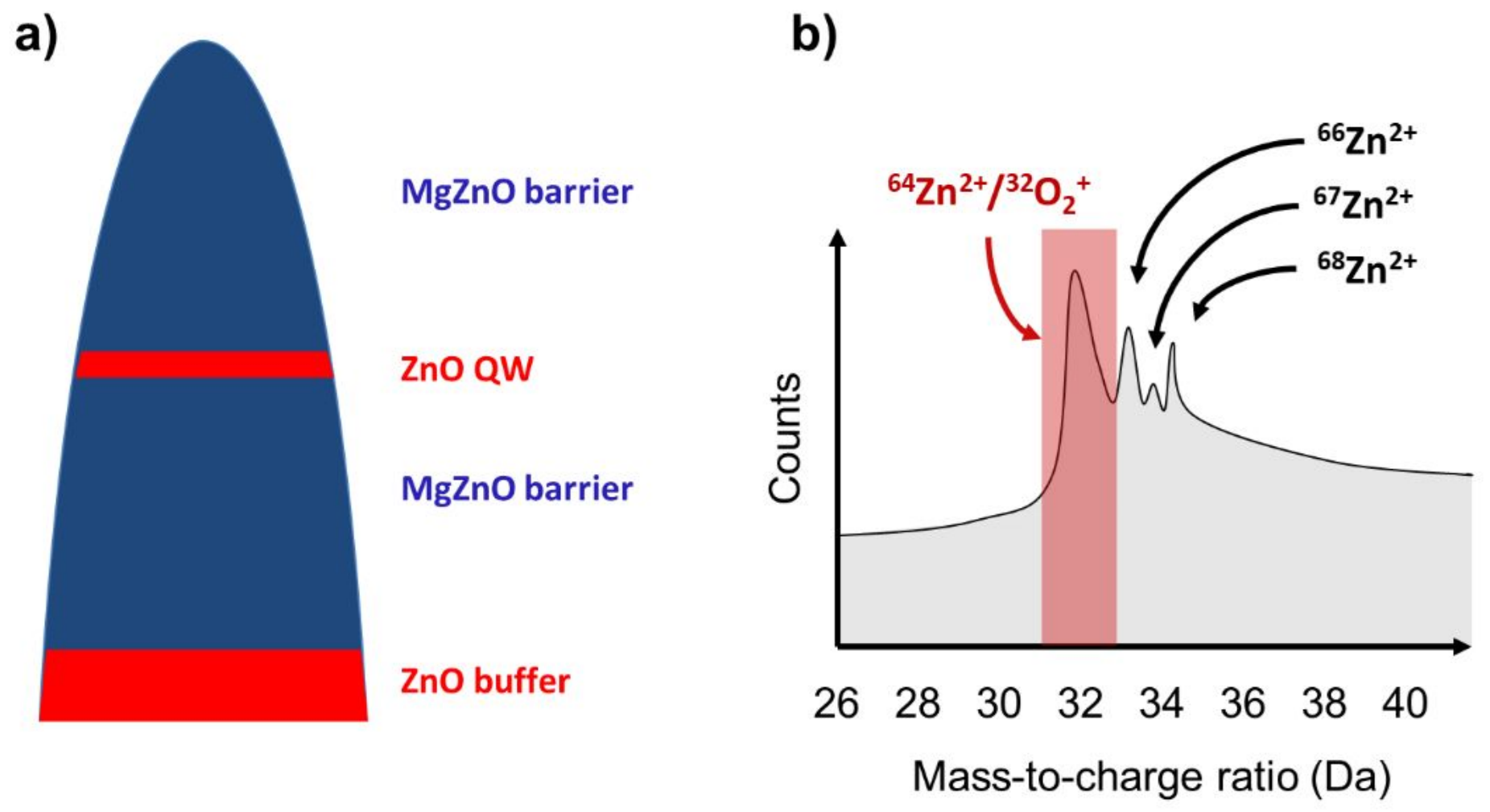

*Figure 11: a) Scheme of a tip sample extracted from a ZnO/$Mg_xZn_{1-x}O$ system; b) Partial mass spectrum around the potential MPO between $^{64}Zn^{2+}$ and $O_2^+$ peaks.*

For this next study, it has been decided to analyze a $ZnO/Mg_xZn_{1-x}O$ quantum well (QW) system, for specifically resolving the potential MPO located at 32 Da between $^{64}Zn^{2+}$ and $O_2^+$ peaks (Amirifar et al., 2015; Dawahre et al., 2011; Park et al., 2013) (Figure 11b). The quantum well system is characterized by a heterostructure made from semiconducting barriers, with large band-gaps, surrounding nanometric layers, also known as quantum wells, with smaller band-gaps (Figure 11a). Through this structure, both electrons and holes can be confined in the central quantum well region, having lower conduction band energy, in order to operate in different applications such as, lasers, photodetectors, modulators, or switches (Anon, 2002). Apart from the exitonic properties of this material, the interest of using the $ZnO/Mg_xZn_{1-x}O$ system is to get a visual evidence of the energy sensitivity of the CF detector. Indeed, looking at the spatial distributions of Zn and O from the conventional APT detector (Figure 12), it can be observed that the oxygen is almost homogenously distributed along the sample, while the zinc is introduced in a higher density at the ZnO QW location. But the key point is that, from the conventional APT detector point of view, it is not possible to extract the spatial distributions of $^{64}Zn^{2+}$ and $^{32}O_2^+$ due to their spectral overlap.

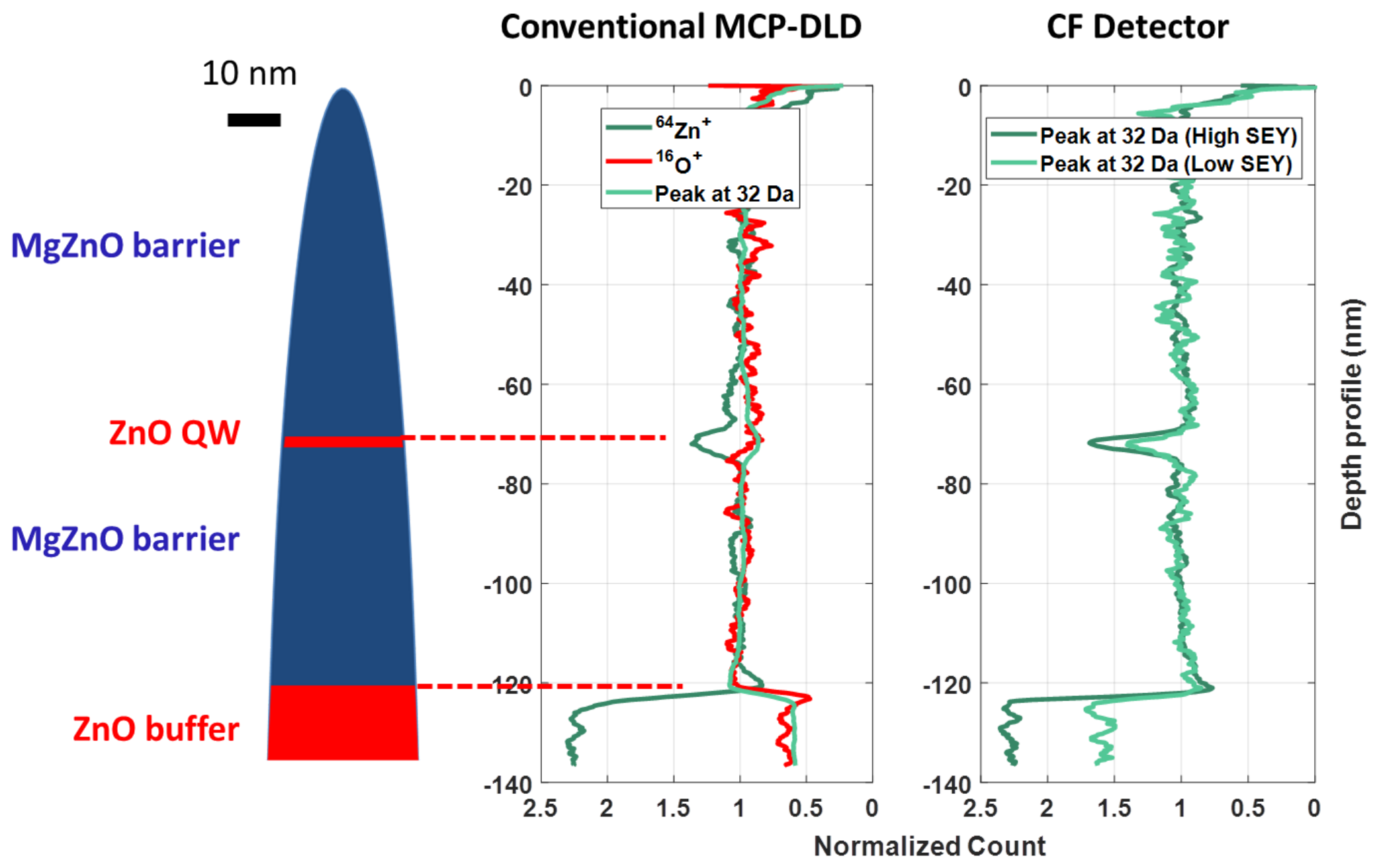

*Figure 12: Normalized depth profiles comparison from the developed detector. The outburst of $^{64}Zn^{2+}$ that could not be observed at the QW position on the conventional*

*MCP-DLD detector can now be observed through the pseudo-energy sensitivity of the CF detector.*

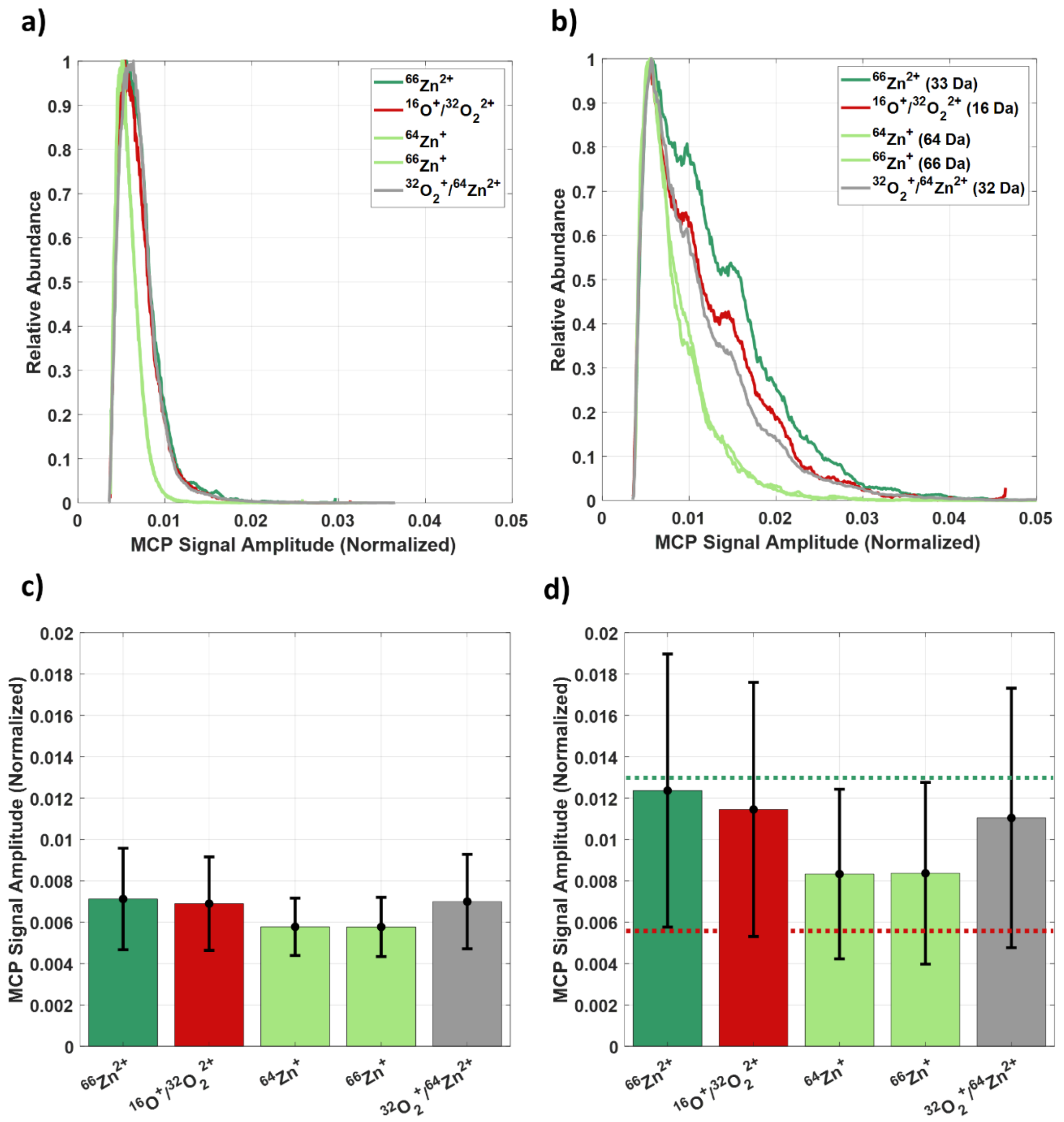

*Figure 13: MCP pulse height distributions originating from the analysis of the ZnO/$Mg_xZn_{1-x}O$ system: a) and c) from "no-foil" area, b) and d) from CF area (at △△△$V_{CF-MCP}$ = 600 V). Histograms from the CF area (b and d) show that detected ions with multiple charge state introduce higher amplitudes than ions with single charge state. With the aim of spatially filtering $^{64}Zn^{2+}$ and $^{32}O_2^{+}$ ions, amplitude levels have been arbitrary set with the help of amplitude standard deviations from single and double charge ions; the red dotted line represents the upper limit of amplitudes*

*attributed to $^{32}O_2^+$ ions and the green dotted line represents the lower limit of amplitudes attributed to $^{64}Zn^{2+}$.*

In the same manner as what has been done for the previous analysis of Fe-Al alloy, a comparison between the two detection areas has been performed through their respective MCP pulse height distribution, normalized with applied DC voltages in the range 8 – 8.4 kV. Figures 13b and 13d introduce the energy sensitivity of the CF detector through the different MCP pulse height distributions introducing higher MCP amplitudes for double charge ions against single charge ions. It can also be observed that the energy resolution of the CF detector is not sufficient for completely discriminating $^{64}Zn^{2+}$ ions from $^{32}O_2^+$ ions from the mass peak at 32 Da. However, considering that $^{32}O_2^+$ ions have theoretically lower SEY than $^{64}Zn^{2+}$ ions, it could be possible to get a first evaluation of the ability of the CF detector to spatially resolve those two elements.

The direct comparison of the depth profiles between the conventional MCP-DLD detector and the CF detector (Figure 12), clearly shows that $^{64}Zn^{2+}$ ions can be partly recovered through the filtering of high MCP amplitudes induced by ions from the mass peak at 32 Da (Figure 13d). However, it can also be observed that the depth profile of the mass peak at 32 Da, filtered with lower MCP amplitudes (lower SEY), seems to follow the same depth profile as Zn isotopes instead of $^{16}O^+$ depth profile. That compositional bias could be explained by a potential selective loss of $^{32}O_2^+$ ions having not enough kinetic energy to induce secondary electrons.

# IV. Conclusion

A first proof of concept of position-energy-sensitive detector has been developed for APT instruments in order to deal with some MPO issues encountered in APT experiments. Through this new type of detector, quantitative and qualitative improvements could be considered for critical materials introducing MPOs.

This new detector based on a thin carbon foil positioned on the front panel of a conventional MCP-DLD detector can generate a number of transmitted SE that mainly depends on both the kinetic energy and the mass of incident particles. Therefore, this study introduces the first experiments on a potential new generation of APT detectors that would be able to resolve MPOs through the energy-sensitivity of thin carbon foils.

In addition to the mass and energy sensitivity of the carbon foil, it has been demonstrated that this type of detector could also be used as a mean for improving the APT detection efficiency. However, it has also been observed that significant losses may occur for massive and low energy ions, caused by their low probability to induce high SEY.

The observed lack of energy-sensitivity of the CF detector requires future in-depth studies to improve its performances. Given that the MCP assembly used in this study introduce a transparency of ~ 60%, it is obvious that one of the first improvements to apply concern the increase of this transparency to reduce the uncertainty on the counting of SEs. That means that the use of existing 90% transparent MCP assemblies could be useful for improving the energy resolution of the CF detector.

The other main improvement could be found in the fine control of the CF thickness, considering that SEs could be generated at different depth, depending on the mass and the energy ranges of analyzed ions. Other future in-depth studies on the Ion Induced Electron Emission phenomenon would be then also necessary to achieve a fine control of this parameter.

## Acknowledgments

This work was financially supported by CAMECA and the University of Rouen. The authors acknowledge Constance Stoner and Robert Stoner from ACF-Metals for assistance with the design and manufacturing of carbon foil mountings and Pr. Lorenzo Rigutti for specimen preparation of MgZnO heterostructures.

## References

ALLEGRINI, F., EBERT, R. W. & FUNSTEN, H. O. (2016). Carbon foils for space plasma instrumentation: CARBON FOILS FOR SPACE. *Journal of Geophysical Research: Space Physics* **121**, 3931–3950.

ALLEGRINI, F., WIMMER-SCHWEINGRUBER, R. F., WURZ, P. & BOCHSLER, P. (2003). Determination of low-energy ion-induced electron yields from thin carbon foils. *Nuclear Instruments* *and Methods in Physics Research Section B: Beam Interactions with Materials and Atoms* **211**, 487–494.

AMIRIFAR, N., LARDÉ, R., TALBOT, E., PAREIGE, P., RIGUTTI, L., MANCINI, L., HOUARD, J., CASTRO, C., SALLET, V., ZEHANI, E., HASSANI, S., SARTEL, C., ZIANI, A. & PORTIER, X. (2015). Quantitative analysis of doped/undoped ZnO nanomaterials using laser assisted atom probe tomography: Influence of the analysis parameters. *Journal of Applied Physics* **118**, 215703.

BACCHI, C. (2020). 'New Generation of Position-Sensitive Detectors for the Development of the Atom Probe Tomography'. Thesis https://tel.archives-ouvertes.fr/tel-03099387.

BACCHI, C., DA COSTA, G. & VURPILLOT, F. (2019). Spatial and Compositional Biases Introduced by Position Sensitive Detection Systems in APT: A Simulation Approach. *Microscopy and Microanalysis* **25**, 418–424.

BARAGIOLA, R. A. (1993a). Principles and mechanisms of ion induced electron emission. *Nuclear Instruments and Methods in Physics Research Section B: Beam Interactions with Materials and Atoms* **78**, 223–238.

——— (1993b). Principles and mechanisms of ion induced electron emission. *Nuclear Instruments and Methods in Physics Research Section B: Beam Interactions with Materials and Atoms* **78**, 223–238.

BARNSTEDT, J. (2016). *Advanced Practical Course Microchannel Plate Detectors*. University of Tübingen.

BAS, P., BOSTEL, A., DECONIHOUT, B. & BLAVETTE, D. (1995). A general protocol for the reconstruction of 3D atom probe data. *Applied Surface Science* **87–88**, 298–304.

BRODERICK, S. R., BRYDEN, A., SURAM, S. K. & RAJAN, K. (2013). Data mining for isotope discrimination in atom probe tomography. *Ultramicroscopy* **132**, 121–128.

BUHR, H., MENDES, M. B., NOVOTNÝ, O., SCHWALM, D., BERG, M. H., BING, D., HEBER, O., KRANTZ, C., ORLOV, D. A., RAPPAPORT, M. L., SORG, T., STÜTZEL, J., VARJU, J., WOLF, A. & ZAJFMAN, D. (2010). Energy-sensitive imaging detector applied to the dissociative recombination of D 2 H +. *Physical Review A* **81**, 062702.

CEREZO, A., GODFREY, T. J. & SMITH, G. D. W. (1988). Application of a position-sensitive detector to atom probe microanalysis. *Review of Scientific Instruments* **59**, 862–866.

Chen, C. Y., Hun, C. W., Chen, S.-F., Chen, C. C., Lin, J. S., Johnson, S. S., Noel, N., Juliely, N. & Luo, Z. (2015). Fabrication of Nanoscale Cesium Iodide (CsI) Scintillators for High-Energy Radiation Detection. *Reviews in Nanoscience and Nanotechnology* **4**, 26–49.

Chianelli, C., Ageron, P., Bouvet, J. P., Karolak, M., Martin, S. & Robert, J. P. (1988). Weakly ionizing charged particle detectors with high efficiency using transitory electronic secondary emission of porous CsI. *Nuclear Instruments and Methods in Physics Research Section A: Accelerators, Spectrometers, Detectors and Associated Equipment* **273**, 245–256.

Da Costa, G., Vurpillot, F., Bostel, A., Bouet, M. & Deconihout, B. (2005). Design of a delay-line position-sensitive detector with improved performance. *Review of Scientific Instruments* **76**, 013304.

Da Costa, G., Wang, H., Duguay, S., Bostel, A., Blavette, D. & Deconihout, B. (2012). Advance in multi-hit detection and quantization in atom probe tomography. *Review of Scientific Instruments* **83**, 123709.

Dawahre, N., Brewer, J., Shen, G., Harris, N., Wilbert, D. S., Butler, L., Balci, S., Baughman, W., Kim, S. M. & Kung, P. (2011). Nanoscale characteristics of single crystal zinc oxide nanowires. In *2011 11th IEEE International Conference on Nanotechnology*, pp. 640–645. Portland, OR, USA: IEEE http://ieeexplore.ieee.org/document/6144626/ (Accessed April 20, 2020).

Dennison, J. R., Sim, A. & Thomson, C. D. (2006). Evolution of the Electron Yield Curves of Insulators as a Function of Impinging Electron Fluence and Energy. *IEEE Transactions on Plasma Science* **34**, 2204–2218.

Drexler, C. G. & DuBois, R. D. (1996). Energy- and angle-differential yields of electron emission from thin carbon foils after fast proton impact. *Physical Review A* **53**, 1630–1637.

Engberg, D. L. J., Johnson, L. J. S., Jensen, J., Thuvander, M. & Hultman, L. (2018). Resolving mass spectral overlaps in atom probe tomography by isotopic substitutions – case of TiSi15N. *Ultramicroscopy* **184**, 51–60.

FUJII, G., UKIBE, M., SHIKI, S. & OHKUBO, M. (2015). Development of Array Detectors with Three-Dimensional Structure toward 1000 Pixels of Superconducting Tunnel Junctions. *IEICE Transactions on Electronics* **E98.C**, 192–195.

FUNSTEN, H. O., RITZAU, S. M., HARPER, R. W. & KORDE, R. (2004). Fundamental limits to detection of low-energy ions using silicon solid-state detectors. *Applied Physics Letters* **84**, 3552–3554.

GAULT, B., HALEY, D., DE GEUSER, F., MOODY, M. P., MARQUIS, E. A., LARSON, D. J. & GEISER, B. P. (2011). Advances in the reconstruction of atom probe tomography data. *Ultramicroscopy* **111**, 448–457.

GLOECKLER, G. & HSIEH, K. C. (1979). Time-of-flight technique for particle identification at energies from 2–400 keV/nucleon. *Nuclear Instruments and Methods* **165**, 537–544.

GORUGANTHU, R. R. & WILSON, W. G. (1984). Relative electron detection efficiency of microchannel plates from 0–3 keV. *Review of Scientific Instruments* **55**, 2030–2033.

HASSELKAMP, D. (1992). Kinetic electron emission from solid surfaces under ion bombardment. In *Particle Induced Electron Emission II*vol. 123, *Springer Tracts in Modern Physics*, Hasselkamp, Dietmar, Rothard, H., Groeneveld, K.-O., Kemmler, J., Varga, P. & Winter, H. (Eds.), pp. 1–95. Berlin, Heidelberg: Springer Berlin Heidelberg http://link.springer.com/10.1007/BFb0038298 (Accessed April 16, 2020).

HATZOGLOU, C., DA COSTA, G. & VURPILLOT, F. (2019). Enhanced dynamic reconstruction for atom probe tomography. *Ultramicroscopy* **197**, 72–82.

HILL, A. G., BUECHNER, W. W., CLARK, J. S. & FISK, J. B. (1939). The Emission of Secondary Electrons Under High Energy Positive Ion Bombardment. *Physical Review* **55**, 463–470.

JACKSON, W. J. (1927). Secondary Emission from Mo Due to Bombardment by High Speed Positive Ions of The Alkali Metals. *Physical Review* **30**, 473–478.

JAGUTZKI, O., MERGEL, V., ULLMANN-PFLEGER, K., SPIELBERGER, L., SPILLMANN, U., DÖRNER, R. & SCHMIDT-BÖCKING, H. (2002). A broad-application microchannel-plate detector system for advanced particle or photon detection tasks: large area imaging, precise multi-hit timing information and high detection rate. *Nuclear Instruments and Methods in Physics Research Section A: Accelerators, Spectrometers, Detectors and Associated Equipment* **477**, 244–249.

KELLY, T. F. (2011). Kinetic-Energy Discrimination for Atom Probe Tomography: Review Article. *Microscopy and Microanalysis* **17**, 1–14.

KIRCHHOFER, R., DIERCKS, D. R., GORMAN, B. P., IHLEFELD, J. F., KOTULA, P. G., SHELTON, C. T. & BRENNECKA, G. L. (2014). Quantifying Compositional Homogeneity in Pb(Zr,Ti)$O_3$ Using Atom Probe Tomography Green, D. J. (Ed.). *Journal of the American Ceramic Society* **97**, 2677–2697.

KUZNETSOV, A. V., VELDHUIZEN, E. J., WESTERBERG, L., LYAPIN, V. G., ALEKLETT, K., LOVELAND, W., BONDORF, J., JAKOBSSON, B., WHITLOW, H. J. & EL BOUANANI, M. (2000). A compact Ultra-High Vacuum (UHV) compatible instrument for time of flight–energy measurements of slow heavy reaction products. *Nuclear Instruments and Methods in Physics Research Section A: Accelerators, Spectrometers, Detectors and Associated Equipment* **452**, 525–532.

LA FONTAINE, A., ZAVGORODNIY, A., LIU, H., ZHENG, R., SWAIN, M. & CAIRNEY, J. (2016). Atomic-scale compositional mapping reveals Mg-rich amorphous calcium phosphate in human dental enamel. *Science Advances* **2**, e1601145.

LADISLAS WIZA, J. (1979). Microchannel plate detectors. *Nuclear Instruments and Methods* **162**, 587–601.

MAIER-KOMOR, P. (1993). Carbon foils for nuclear accelerator experiments. *Nuclear Instruments and Methods in Physics Research Section B: Beam Interactions with Materials and Atoms* **79**, 841–844.

MARCEAU, R. K. W., CEGUERRA, A. V., BREEN, A. J., RAABE, D. & RINGER, S. P. (2015). Quantitative chemical-structure evaluation using atom probe tomography: Short-range order analysis of Fe–Al. *Ultramicroscopy* **157**, 12–20.

MARTIN, A. J., WEI, Y. & SCHOLZE, A. (2018). Analyzing the channel dopant profile in next-generation FinFETs via atom probe tomography. *Ultramicroscopy* **186**, 104–111.

MEISENKOTHEN, F., STEEL, E. B., PROSA, T. J., HENRY, K. T. & PRAKASH KOLLI, R. (2015). Effects of detector dead-time on quantitative analyses involving boron and multi-hit detection events in atom probe tomography. *Ultramicroscopy* **159**, 101–111.

MILLER, M. K. (1987). THE EFFECTS OF LOCAL MAGNIFICATION AND TRAJECTORY ABERRATIONS ON ATOM PROBE ANALYSIS. *Le Journal de Physique Colloques* **48**, C6-565-C6-570.

MILLER, M. K. & FORBES, R. G. (2014). *Atom-Probe Tomography*. Boston, MA: Springer US http://link.springer.com/10.1007/978-1-4899-7430-3 (Accessed July 9, 2018).

MIODOWNIK, M. (2015). Materials for the 21st century: What will we dream up next? *MRS Bulletin* **40**, 1188–1197.

MONTAGNOLI, G., STEFANINI, A. M., TROTTA, M., BEGHINI, S., BETTINI, M., SCARLASSARA, F., SCHIAVON, V., CORRADI, L., BEHERA, B. R., FIORETTO, E., GADEA, A., LATINA, A., SZILNER, S., DONÀ, L., RIGATO, M., KONDRATIEV, N. A., CHIZHOV, A. YU., KNIAJEVA, G., KOZULIN, E. M., POKROVSKIY, I. V., VOSKRESSENSKY, V. M. & ACKERMANN, D. (2005). The large-area micro-channel plate entrance detector of the heavy-ion magnetic spectrometer PRISMA. *Nuclear Instruments and Methods in Physics Research Section A: Accelerators, Spectrometers, Detectors and Associated Equipment* **547**, 455–463.

MÜLLER, A., DJURIĆ, N., DUNN, G. H. & BELIĆ, D. S. (1986). Absolute detection efficiencies of microchannel plates for 0.1–2.3 keV electrons and 2.1–4.4 keV $Mg^{+}$ ions. *Review of Scientific Instruments* **57**, 349–353.

MURDOCK, J. W. & MILLER, G. H. (1995). *Secondary electron emission due to positive ion bombardment*. https://lib.dr.iastate.edu/ameslab_iscreports/106.

OHKUBO, M., SHIGETOMO, S., UKIBE, M., FUJII, G. & MATSUBAYASHI, N. (2014). Superconducting Tunnel Junction Detectors for Analytical Sciences. *IEEE Transactions on Applied Superconductivity* **24**, 1–8.

OLIPHANT, M. L. E. (1930). The Liberation of Electrons from Metal Surfaces by Positive Ions. Part I. Experimental. *Proceedings of the Royal Society A: Mathematical, Physical and Engineering Sciences* **127**, 373–387.

PALMBERG, P. W. (1967). Secondary Emission Studies on Ge and Na-Covered Ge. *Journal of Applied Physics* **38**, 2137–2147.

PARK, S., JUNG, W. & PARK, C. (2013). Distribution of nickel in zinc oxide nanowalls on sapphire substrate investigated using atom probe tomography. *Scripta Materialia* **68**, 1000–1003.

PENG, Z., VURPILLOT, F., CHOI, P.-P., LI, Y., RAABE, D. & GAULT, B. (2018). On the detection of multiple events in atom probe tomography. *Ultramicroscopy* **189**, 54–60.

*Physics of Quantum Well Devices* (2002).Dordrecht: Kluwer Academic Publishers http://link.springer.com/10.1007/0-306-47127-2 (Accessed April 20, 2020).

RITZAU, S. M. & BARAGIOLA, R. A. (1998). Electron emission from carbon foils induced by keV ions. *Physical Review B* **58**, 2529–2538.

ROTHARD, H., CARABY, C., CASSIMI, A., GERVAIS, B., GRANDIN, J.-P., JARDIN, P., JUNG, M., BILLEBAUD, A., CHEVALLIER, M., GROENEVELD, K.-O. & MAIER, R. (1995). Target-thickness-dependent electron emission from carbon foils bombarded with swift highly charged heavy ions. *Physical Review A* **51**, 3066–3078.

RYMZHANOV, R. A., MEDVEDEV, N. A. & VOLKOV, A. E. (2015). Electron emission from silicon and germanium after swift heavy ion impact: Electron emission from Si and Ge after swift heavy ion impact. *physica status solidi (b)* **252**, 159–164.

ŠARO, Š., JANIK, R., HOFMANN, S., FOLGER, H., HESSBERGER, F. P., NINOV, V., SCHÖTT, H. J., KABACHENKO, A. P., POPEKO, A. G. & YEREMIN, A. V. (1996a). Large size foil-microchannel plate timing detectors. *Nuclear Instruments and Methods in Physics Research Section A: Accelerators, Spectrometers, Detectors and Associated Equipment* **381**, 520–526.

——— (1996b). Large size foil-microchannel plate timing detectors. *Nuclear Instruments and Methods in Physics Research Section A: Accelerators, Spectrometers, Detectors and Associated Equipment* **381**, 520–526.

SEOL, J.-B., RAABE, D., CHOI, P., PARK, H.-S., KWAK, J.-H. & PARK, C.-G. (2013). Direct evidence for the formation of ordered carbides in a ferrite-based low-density Fe–Mn–Al–C alloy studied by transmission electron microscopy and atom probe tomography. *Scripta Materialia* **68**, 348–353.

SHAPIRA, D., LEWIS, T. A. & HULETT, L. D. (2000). A fast and accurate position-sensitive timing detector based on secondary electron emission. *Nuclear Instruments and Methods in Physics Research Section A: Accelerators, Spectrometers, Detectors and Associated Equipment* **454**, 409–420.

THUVANDER, M., WEIDOW, J., ANGSERYD, J., FALK, L. K. L., LIU, F., SONESTEDT, M., STILLER, K. & ANDRÉN, H.-O. (2011). Quantitative atom probe analysis of carbides. *Ultramicroscopy* **111**, 604–608.

TÖGLHOFER, K., AUMAYR, F. & WINTER, H. P. (1993). Ion-induced electron emission from metal surfaces — insights from the emission statistics. *Surface Science* **281**, 143–152.

TOMITA, S., YODA, S., UCHIYAMA, R., ISHII, S., SASA, K., KANEKO, T. & KUDO, H. (2006). Nonadditivity of convoy- and secondary-electron yields in the forward-electron emission from thin carbon foils under irradiation of fast carbon-cluster ions. *Physical Review A* **73**, 060901.

TSONG, T. T. (1990). *Atom-probe field ion microscopy: Field ion emission and surfaces and interfaces at atomic resolution*. Cambridge: Cambridge University Press http://ebooks.cambridge.org/ref/id/CBO9780511599842 (Accessed July 9, 2018).

VURPILLOT, F. (2016). Field Ion Emission Mechanisms. In *Atom Probe Tomography*, pp. 17–72. Elsevier http://linkinghub.elsevier.com/retrieve/pii/B9780128046470000024 (Accessed July 9, 2018).

VURPILLOT, F., BOSTEL, A. & BLAVETTE, D. (2000). Trajectory overlaps and local magnification in three-dimensional atom probe. *Applied Physics Letters* **76**, 3127–3129.

WARD, B. W., SHAVER, D. C. & WARD, M. L. (1985). Repair Of Photomasks With Focussed Ion Beams. , Blais, P. D. (Ed.), p. 110. Santa Clara http://proceedings.spiedigitallibrary.org/proceeding.aspx?doi=10.1117/12.947491 (Accessed January 30, 2020).

WEATHERS, L. W. & TSANG, M. B. (1996). Fabrication of thin scintillator foils. *Nuclear Instruments and Methods in Physics Research Section A: Accelerators, Spectrometers, Detectors and Associated Equipment* **381**, 567–568.

WORTMANN, M., LUDWIG, A., MEIJER, J., REUTER, D. & WIECK, A. D. (2013). High-resolution mass spectrometer for liquid metal ion sources. *Review of Scientific Instruments* **84**, 093305.

ZIEGLER, J. F. & BIERSACK, J. P. (1985). The Stopping and Range of Ions in Matter. In *Treatise on Heavy-Ion Science: Volume 6: Astrophysics, Chemistry, and Condensed Matter*, Bromley, D. A. (Ed.), pp. 93–129. Boston, MA: Springer US https://doi.org/10.1007/978-1-4615-8103-1_3 (Accessed January 30, 2020).

ZIEGLER, J. F., ZIEGLER, M. D. & BIERSACK, J. P. (2010). SRIM – The stopping and range of ions in matter (2010). *Nuclear Instruments and Methods in Physics Research Section B: Beam Interactions with Materials and Atoms* **268**, 1818–1823.